\documentclass[12pt]{article}

\usepackage{graphicx}

\newcommand {\be}{\begin{equation}} 
\newcommand{\ee}{\end{equation}}    

\def\dds1{\frac{\partial}{\partial s_1}}

\def\d{d\kern-0.8 ex\vrule height 1.3 ex depth-1.24 ex width 0.7 ex
\kern 0.15 ex}
\def\D{D\kern-1.7 ex\vrule height .87 ex depth-0.8 ex width 0.7 ex
\kern 0.95 ex}

\evensidemargin
0.3cm
\begin{document}
\baselineskip 20 pt

\begin{center}

\Large{\bf On quantum plasma: a plea for a common sense }

\end{center}

\vspace{0.7cm}

\begin{center}

{\bf  J. Vranjes$^1$, B. P. Pandey$^2$, and S. Poedts$^3$}

{\em $^{1}$ Von Karman Institute, Waterloosesteenweg 72, 1640 Sint-Genesius-Rode, Brussels,
 Belgium.
}

\vspace{5mm}

  {\em  $^{2}$Department of Physics, Macquarie University, Sydney, NSW
2109, Australia.}
\vspace{5mm}

{\em $^{3}$Center for Plasma Astrophysics,  and Leuven Mathematical Modeling and Computational Science Center
(LMCC), K. U. Leuven, Celestijnenlaan 200B, 3001 Leuven,  Belgium.
}

\vspace{5mm}

\end{center}
\vspace{2cm}

{\bf Abstract:} The quantum plasma theory has flourished in the past  few years without much regard to
 the physical validity of the formulation or its connection to any real physical system. It is argued
 here that there is a very limited physical ground for the application of such a theory.

\vspace{2cm}


\pagebreak

In the past a few years one could observe  a lot of activity in the literature regarding the quantum plasma. The amount
of publications  is already measured  by three-digit numbers.  Practically without exceptions those works deal with
collective interactions in plasmas and with consequent waves, instabilities, soliton formations, vortices, etc., and
also in almost all cases such studies are performed within the framework of fluid theory. It is a common wisdom that
the fluid approach is valid only if the spatial scales of the physical problem are considerably larger than the mean
distance between the plasma particles $\lambda_m$, and also, above the mean free path of particles
$\lambda_f=v_{tj}/\nu_j$, $v_{tj}^2= \kappa T_j/m_j$. The latter applies to the kinetic theory too, which operates with
distribution functions (Maxwellian or some of its modifications), and those are mainly achieved in the processes of
thermalization (i.e.\ collisions) or fluctuations.   At such scales the quantum effects are simply out of scope.

 Normally, the quantum effects would appear primarily for lighter particles (electrons).  In reality however,
  the quantum effects in plasmas are typically of very limited importance. This may be seen from basic books on
  plasma theory, see e.g.\ in Ref.~[1].   The electron temperatures above which the classical plasma description
  is valid are determined by the following limit
\be
T_e (n_e)=\left[\left(\frac{\hbar}{2}\right)^3 \left(\frac{1}{3 m_e \kappa}\right)^{3/2} \cdot  n_e\right]^{2/3}=7.37 \cdot 10^{-17} n_e^{2/3}. \label{e1}
\ee
Here,   the electron number density is per cubic meter, and the electron temperature is in
Kelvin degrees.    The temperature-density boundary  separating
the classic and quantum domains, for some typical plasma densities,  is presented in Fig.~1. It is seen that extremely
low temperatures  (on the order of $10^{-5}$ K) are needed in order to observe quantum effects.

\begin{figure}
\includegraphics[height=6.5cm, bb=14 14 295 235, clip=]{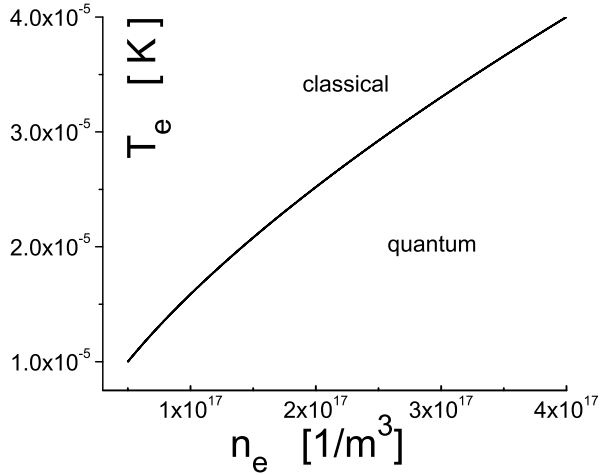}
\vspace*{-5mm} \caption{ The locus of pairs $n_e, T_e$ determining the boundary between the classical and quantum
domains. }\label{fig1}
\end{figure}

This is presented also in Table~1 for a much larger span of the electron number density, see the first two rows. For
electrons, some quantum effects may become of interest only at the number density of the order of $10^{30}\;$m$^{-3}$
and small temperatures of a few thousand K. In this limit only, the uncertainty in the position becomes comparable to
the distance between electrons.
However, for ions even the number densities of around $10^{30}\;$m$^{-3}$ are not high enough. This is shown by several
values in brackets in Table~1.
It is obvious that the quantum domain  is never of any importance in most plasmas, except in the core of neutron stars
which can be characterized as a dense ($n\sim  10^{35}-10^{41}\;$m$^{-3}$) cold plasma  \cite{vv}.

Because the mean distance between particles is $\lambda_m \simeq 1/n^{1/3}$, for those extremely large number densities
of the order of $n_j\simeq 10^{30}-10^{40}\;$m$^{-3}$ one finds $\lambda_m\simeq 10^{-10}-10^{-13}\;$m. Hence, the
 quantum effects are expected at atomic and subatomic scales, i.e.,  those on the order of the Bohr radius or below it \cite{kr,r2}.
Studying quantum corrections therefore may have sense for elementary processes like the  ionization, attachment, detachment etc.

\begin{table*}
\caption{The electron number density and the corresponding temperature Eq.~(\ref{e1}) below  which the quantum effects should be included (the values in brackets
are  for protons), and the corresponding ratio of the mean distance between particles  and the Debye length. } \label{table:2} \centering
\vspace{0.3cm}
\begin{tabular}{ c c c  c c c c c c    }     

\hline
\\
$n_e$ [m$^{-3}$ ] & $10^6 $ & $ 10^{12}$ & $ 10^{18}$ & $ 10^{24}$ & $ 10^{26}$ & $ 10^{28}$ & $ 10^{29}$ & $ 10^{30}$   \\
$T_e$ [K]  & $7\cdot 10^{-13} $ & $7\cdot 10^{-9} $ & $7\cdot 10^{-5} $ & $ 0.7$ & $ 16$ & $342(0.18)$ & $ 1587(0.85)$ & $7369  (3.9)$   \\
  $\lambda_m/r_{de}$  &  &   &  & $ 173$ & $ 78$ & $36$ & $ 25$ & $17$   \\
\hline
\end{tabular}

\end{table*}

 We stress also that the kinetic and fluid theories, used  in studying  collective interactions in plasmas and therefore
 the phenomena mentioned above (waves, instabilities, etc), imply
weakly non-ideal plasmas. Those are plasmas close to thermodynamic equilibrium, and in the same time  comprising particles
 whose energy of thermal motion is considerably above the energy of the electrostatic interaction of the charged particles.
 In practical terms this implies the presence of a relatively high  number of particles within  the Debye sphere, or in other
 words the mean distance  $\lambda_m$ between two charged particles should be well below the Debye
 radius $r_{dj}= v_{tj}/\omega_{pj}$, $\omega_{pj}=[q_j^2 n_0/(\varepsilon_0 m_j)]^{1/2}$.  The total internal energy of
 such a plasma can be written
as $U=U_{th}+ U_{int}$ where the first and second terms are, respectively,  the thermal energy and the energy of electrostatic
 interaction. This can further  be written as
  $U=(3 n \kappa T/2 ) [1- \lambda_m^3/(12 \pi r_d^3)]$, where $n=\sum n_j$. Here, the ratio $\lambda_m^3/r_d^3$ must be well
  below unity (the standard condition of a weakly non-ideal plasma).
    However, the third row in Table~1 shows just  the opposite in the domain of temperatures and densities where the quantum
    effects do play a role: the mean distance between the particles is far above the Debye radius. The plasma is thus far
     from total or partial thermodynamic equilibrium. This thermodynamic equilibrium is in fact an underlaying condition for
      the {\it collective behavior} of a standard  plasma description used in numerous papers. Since this condition is
      often violated, the application of such a description is not valid.

  The degeneracy of electrons, as an additional manifestation of quantum effects,  appears at temperatures (in K) below
\[ T_{deg}=[h^2/(2 m_e \kappa)] [10^{-6}  n_e/(2.76 \pi)]^{3/2}.
\]
  It is seen that at $n_{e}=10^{30}\;$m$^{-3}$ this value for electrons  becomes very low,  viz.\ $T_{deg}=41\;$K. Such
  temperatures can indeed be found in astrophysical molecular clouds, however, the accompanying plasma  density is many
  (e.g. 15-20) orders  of magnitude below the necessary value. Hence,  the degeneracy is of no importance.

   To summarize,  our criticism is aimed at the huge number of studies on waves  and instabilities in plasmas,
   particularly  within the fluid domain, that is so frequently seen in the literature in the past a few years.
Typically, in these studies some 'quantum' correction terms are included in the momentum equations for electrons
(and ions as well), and this all for parameters where they can not possibly play any role. Particularly meaningless
are such examples  of  works dealing with wave dynamics in dusty plasmas, where, first, spatial scales at which dust
 grain dynamics takes place are many orders  of magnitude above any quantum scale correction, and, second, it is
 clear that finding an appropriate physical system (extremely dense and simultaneously extremely cold) to apply
  these quantum corrections  becomes practically impossible. In our view, the arguments given in the text
    above should be kept in mind and the criteria for the applicability of the theory should be checked before any practical step.



\end{document}